\documentclass[final,3p,times,2024]{TeXStyle/elsarticle}
 \usepackage{pdfpages}
 \usepackage{graphicx}
 \usepackage[T1]{fontenc}

 \usepackage{enumitem}
 \usepackage{subfiles}   
 \usepackage{amsmath,xfrac}
 \usepackage{amssymb}  
 \usepackage{multirow}
 \usepackage{float}
 \usepackage{dblfloatfix}
 \usepackage{bookmark}
 \usepackage{lineno}        
 \usepackage{siunitx} 
 \usepackage{nth}
 \usepackage{url}
 \usepackage{longtable} 
 \usepackage[ruled,vlined]{algorithm2e}
 \usepackage{xcolor}  
 \usepackage[flushleft]{threeparttable}
 \usepackage{caption,nameref}
 \usepackage{subcaption}
 \usepackage{diagbox}
 \usepackage{makecell}
 \usepackage{afterpage}
 \usepackage{newunicodechar}
 \usepackage{hyperref}
 \usepackage{dcolumn} 
 \usepackage{booktabs}
 \usepackage{tabularx} 
 \usepackage{bm, comment} 
 \usepackage{natbib}  
  \usepackage[utf8]{inputenc}

 \newunicodechar{，}{, }

 \setcitestyle{square}
 \biboptions{sort&compress} 
 \usepackage{setspace} 
 \journal{Journal}

\begin{document}
 \begin{frontmatter}
  \title{
 Smart Sensing Breaks the Accuracy Barrier in Battery State Monitoring
 }

\author[1]{Xiaolei~Bian}
\ead{xiaoleib@chalmers.se}
\author[1]{Changfu~Zou\textsuperscript{*}}
\ead{changfu.zou@chalmers.se}
\author[2]{Björn~Fridholm}
\ead{bjorn.fridholm@volvocars.com}
\author[3]{Christian~Sundvall}
\ead{christian.sundvall@novoenergy.se}
\author[1]{Torsten~Wik\textsuperscript{*}}
\ead{tw@chalmers.se}

\cortext[cor1]{Corresponding authors: \texttt{Changfu Zou, Torsten Wik}} 

 \address[1]{Department of Electrical Engineering, Chalmers University of Technology, Gothenburg, 41296, Sweden}
 \address[2]{Volvo Car Corporation, Gunnar Engellaus väg 8, Gothenburg, 40531, Sweden}
 \address[3]{NOVO Energy AB, Gamla Sörredsvägen 26, Gothenburg, 41878, Sweden}

\begin{abstract}
Accurate state-of-charge (SOC) estimation is essential for optimizing battery performance, ensuring safety, and maximizing economic value. Conventional current and voltage measurements, however, have inherent limitations in fully inferring the multiphysics-resolved dynamics inside battery cells. This creates an accuracy barrier that constrains battery usage and reduces cost-competitiveness and sustainability, across industries dependent on battery technology. In this work, we introduce an integrated sensor framework that combines novel mechanical, thermal, gas, optical, and electrical sensors with traditional measurements to break through this barrier. We generate three unique datasets with eleven measurement types and propose an explainable machine-learning approach for SOC estimation. This approach renders the measured signals and the predictive result of machine learning physically interpretable with respect to battery SOC, offering fundamental insights into the time-varying importance of different signals. Our experimental results reveal a marked increase in SOC estimation accuracy--enhanced from 46.1\% to 74.5\%--compared to conventional methods. This approach not only advances SOC monitoring precision but also establishes a foundation for monitoring additional battery states to further improve safety, extend lifespan, and facilitate fast charging.

\end{abstract}

\end{frontmatter}

\section{Introduction}
Lithium-ion (Li-ion) batteries have become indispensable across a spectrum of high-impact applications, spanning electric vehicles, renewable energy storage, and portable electronics \cite{lu2023deep, sulzer2021challenge, severson2019data}. While electromobility and smart grids drive the clean energy transition, portable electronics sustain the quality and continuity of modern life. Despite the widespread use of batteries, many significant challenges remain, such as safety concerns, premature aging, and low efficiency, all of which constrain their cost-benefit and sustainability \cite{jones2022impedance, birkl2017degradation}. 
Addressing these multifaceted challenges requires the development of sophisticated battery management systems (BMS) capable of high-fidelity, real-time monitoring and predictive analytics of battery states. 

Among all key battery states, the state of charge (SOC) is fundamental for achieving various BMS functionalities \cite{andre2013advanced, plett2004extended, lopetegi2024new}, including charge-discharge control, cell balancing \cite{habib2023lithium}, estimation of energy \cite{he2015novel, hao2024novel} and health states \cite{wang2024physics, zhang2023state}, and prediction of power capability \cite{wik2015implementation, guo2023electric} and lifetime \cite{thelen2024probabilistic}. Physically, the SOC of a battery cell reflects the average Li-ion concentration within anode particles, a quantity that cannot be directly measured. Research on SOC estimation began in 1993, shortly after Sony's 1991 invention of Li-ion batteries, with publications in this field growing exponentially to date (see Supplementary Materials Fig.~1), underscoring its popularity and significance. 
As claimed by a magnitude of sophisticated methods in the state of the art, the estimation accuracy has progressed steadily, resulting in average errors now between 0.4\% and 1.1\% \cite{selvaraj2023comprehensive}. However, often plagued by applicability and practicability issues, these advanced methods are rarely implemented into real-world battery usage, putting the state of the industry far behind. SOC estimators typically yield a maximum error of 3--5\%, influenced by factors such as battery chemistry, usage profiles, environmental conditions, and sensor precision \cite{selvaraj2023comprehensive, hannan2017review}. This limited accuracy constrains the usable SOC range to 90--94\% to avoid overcharging or deep discharge. With the global Li-ion battery market projected to reach \$400 billion by 2030  \cite{fleischmann2023battery}, this 6--10\% capacity limitation simply translates to an economic loss of USD\$24--40 billion. Furthermore, such low capacity utilization can significantly hinder the sustainability of batteries, posing challenges for the global green energy transition. 

Accurately monitoring SOC is a fundamentally challenging task due to complex multiphysics inherently involved in a battery cell \cite{liu2024advances}. During battery operation, coupled electrochemical, thermal, and mechanical dynamics are triggered in multiple domains and timescales \cite{reniers2019review}, all of which can affect the value of SOC. 
In the short term, SOC is influenced by charge transfer rates at the electrode surface, lithium-ion diffusion within electrodes, and polarization effects at the electrode-electrolyte interface \cite{klett2012quantifying}. Rapid charge transfer enables quicker SOC adjustments, while slower diffusion limits dynamic response and SOC equilibrium. Additionally, internal resistance and associated thermal effects induce transient voltage changes, affecting SOC estimation accuracy. Long-term SOC estimation is further complicated by degradation processes, including active material breakdown, solid-electrolyte interphase (SEI) layer growth \cite{zhou2020real}, lithium plating on the anode, electrolyte decomposition, and structural changes in electrode materials \cite{an2019scalable}. These collectively reduce battery capacity and increase internal resistance, requiring regular calibration to maintain SOC accuracy. Spatial heterogeneities, such as phase variations within electrodes \cite{matras2022emerging} and thermal gradients \cite{li2023effect}, add additional complexity by causing non-uniform lithium distributions. Together, these processes, compounded by ambient environment fluctuations and measurement noise \cite{li2017state}, complicate the task of SOC estimation.

Due to these challenges and its great importance, many methods have been proposed for SOC estimation over the past two decades, broadly classified into empirical, model-based, and data-driven approaches~\cite{wang2020comprehensive, hu2019state}. Empirical methods rely on predefined relationships and historical patterns~\cite{plett2015battery}; model-based methods use either equivalent circuit or electrochemical models to capture battery dynamics~\cite{chemali2018state, wu2024physics}; and data-driven techniques leverage machine learning to predict SOC from historical and real-time data~\cite{lipu2020data, yang2023state}. Recently, innovative machine learning methods that integrate physical knowledge~\cite{wang2024physical} and cross-domain transfer learning~\cite{yang2024deep, moosavi2024transductive} have emerged as promising strategies for enhancing SOC estimation accuracy. Physics-informed neural networks (PINNs), which explicitly incorporate electrochemical principles, significantly improve model robustness across diverse cell chemistries and operating conditions~\cite{huang2024minn, wang2024physics, wang2025physics}. For instance, Tian \textit{et al.} demonstrated that a physics-informed deep learning method markedly improved SOC prediction, reducing the root mean square error and maximum absolute error by 30.9\% and 64.9\%, respectively~\cite{tian2022battery}. Additionally, transfer learning approaches allow pre-trained models to adapt efficiently to new battery types and cycling conditions using minimal additional data, greatly reducing training time and data requirements~\cite{yang2024deep, moosavi2024transductive}. 
These methods have contributed valuable insights, leading to incremental advances in SOC estimation. 
For these existing methods, current and voltage are the primary inputs  \cite{xing2014state}. 
Yet, as discussed, SOC is influenced by a variety of internal and external factors, meaning that current and voltage signals may inherently not contain sufficient information to fully infer SOC evolution.
This limitation can create a ``glass ceiling'', regardless of the method employed.

Incorporating additional sensors beyond current and voltage could provide crucial insights for fully characterizing battery dynamics.
Researchers are increasingly exploring novel sensing techniques for battery management.
For example, impedance spectroscopy offers valuable information on internal resistance, revealing degradation patterns and enabling predictions of future battery performance \cite{jones2022impedance}.
Cell expansion, which strongly correlates with battery aging, allow for capacity estimation over a narrower SOC range than electrical signals \cite{mohtat2022comparison}.
Force sensors have been employed to detect internal lithium plating, potentially facilitating lithium-plating free charging protocols \cite{huang2022onboard}. 
Acoustic sensors provide real-time monitoring of structural changes within the battery, making it possible to early detect failures \cite{wang2024progress, xiong2024advancing}.
Additionally, fiber optic sensors can characterize internal material properties, offering deeper insights into chemical and electrochemical reactions, such as the oxidation and reduction of active materials  \cite{hedman2020fibre, hedman2021fiber, hedman2023fiber}.
These advancements are all driving the field of battery management forward. 
However, a significant gap remains in what sensors should be deployed and how to integrate their signals for achieving the optimal performance.  

In this work, we propose an integrated sensor framework that leverages novel mechanical, thermal, gas, optical, and electrical sensors with traditional measurements to fundamentally improve and analyze Li-ion battery SOC estimation. Using battery cells of varying chemistries and under diverse conditions, we collect three unique datasets comprising eleven types of measurements. Building upon these datasets, we propose an explainable machine-learning approach, which makes the input signals and output predictions physically interpretable. 
Our experimental results show a significant improvement in SOC estimation accuracy over traditional methods. Specifically, the inclusion of expansion and surface temperature signals increases accuracy by 74.5\%, the addition of optical signals improves accuracy by 46.1\%, and the integration of battery force and anode potential signals contributes to 60.6\% higher accuracy. 

\section{Results}
\label{sec:Results}
In this study, we investigate the SOC estimation of three types of cells using a dual-layer long short-term memory (LSTM) model (see \nameref{sec:Methods}). A total of eleven signals are recorded using different sensors, and three novel signal combinations are explored: cell expansion and surface temperature, internal optical signals, and cell force coupled with electrode potential. The following sections present detailed results for each of these signal combinations, demonstrating their impact on SOC estimation.

\subsection{\textbf{Cell expansion and surface temperature signals} }
\label{sec:cell-expansion}

\begin{figure}[ht]
    \centering
    \includegraphics[trim=0cm 2.2cm 0cm 0cm, clip, width=1\textwidth]{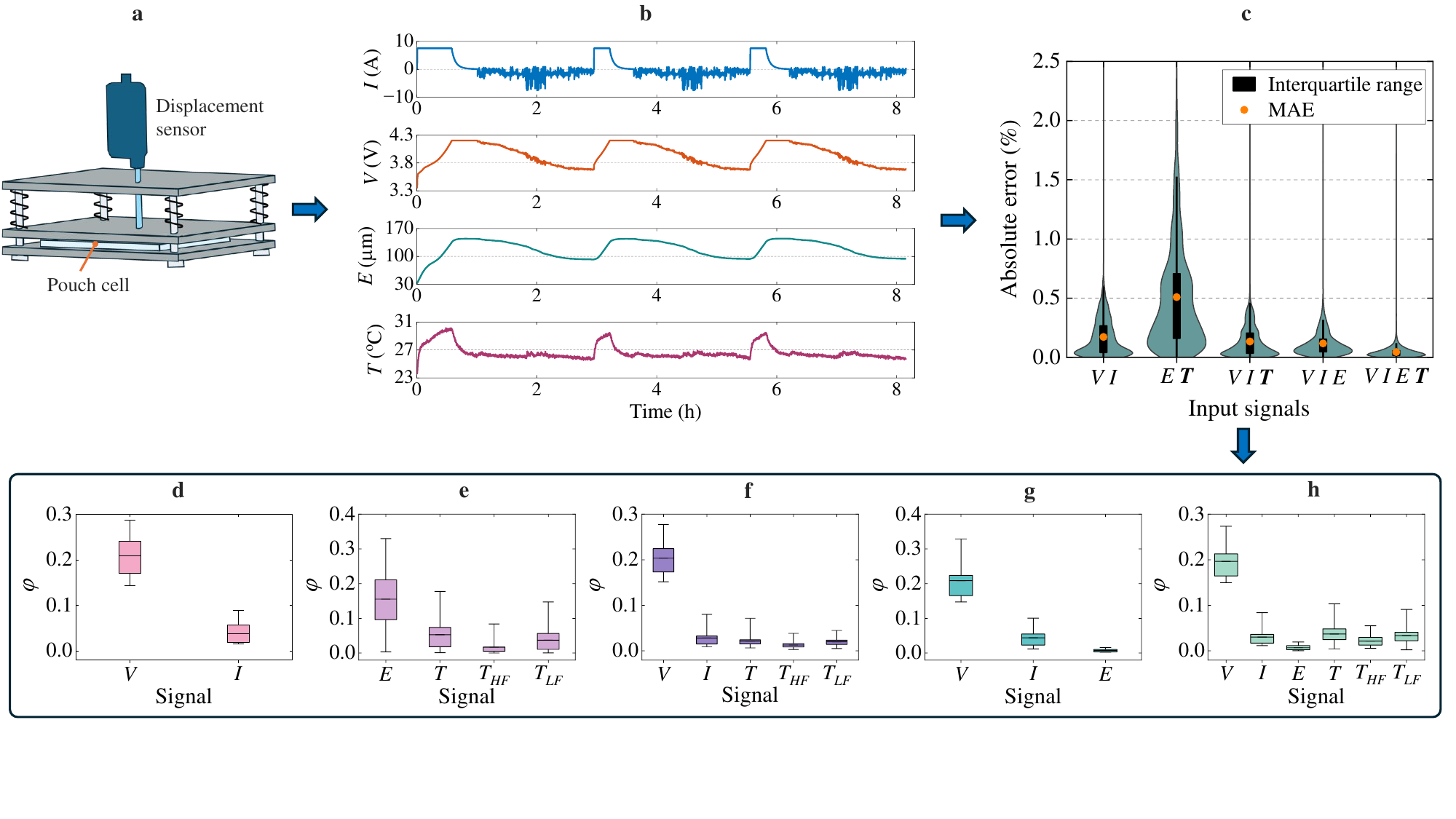}
    \caption{\textbf{Schematic representation and major results for the cell with displacement sensor.} The studied signals include  cell expansion (\textit{E}), and conventional measurements, i.e., voltage (\textit{V}), current (\textit{I}), and surface temperature (\textit{T}). \textbf{a} Illustration of the test apparatus for a cell. \textbf{b} Three cycles of recorded signal data. \textbf{c} SOC estimation results using different signal combinations (the statistical results presented here are steady-state errors, i.e., with the estimation error falling within 5\%). \textbf{d-h} Sensitivity analysis highlighting the significance and contributions of different signals across five scenarios. Here, \textbf{\textit{T}} represents a composite of three temperatures—surface temperature (\textit{T}), high-frequency component ($T_{HF}$), and low-frequency component ($T_{LF}$) (see Supplementary Materials Fig.~3).}
    \label{fig:1_expansionMajor}
\end{figure}

As shown in Fig.~\ref{fig:1_expansionMajor}\textbf{b}, the measured signals include current (\textit{I}), voltage (\textit{V}), cell expansion (\textit{E}), and surface temperature (\textit{T}) (see Supplementary Materials Fig.~2 for details). We found that the inclusion of additional signals can significantly reduce the estimation error. Specifically, as shown in Fig.~\ref{fig:1_expansionMajor}\textbf{c}, the mean absolute error (MAE) for SOC estimation using the conventional signals of current and voltage is 0.17\%. By comparison, when conventional \textit{I-V} signals are excluded, using only \textit{E-\textbf{T}} signals achieves an MAE of 0.51\%. Here, \textbf{\textit{T}} denotes a vector that includes surface temperature (\textit{T}), and its high-frequency component ($T_{HF}$), and low-frequency component ($T_{LF}$) (see Supplementary Materials Fig.~3 for details). Although this accuracy is lower than that of the \textit{I-V} combination, it illustrates a strong correlation of these alternative signals with cell SOC. More importantly, when temperature and expansion are combined with current and voltage, the MAE is reduced to 0.04\% (see Supplementary Materials Fig.~8 for detailed results), representing a 74.5\% improvement in accuracy over the \textit{I-V} only approach. This reveals a promising pathway for achieving a step change, rather than incremental improvements, in SOC estimation.

In addition, our study reveals that decomposing surface temperature (\textit{T}) into high-frequency
($T_{HF}$) and low-frequency ($T_{LF}$) components can enhance the accuracy performance of the machine learning model. 
This approach reduces the MAE by 13.7\%, 24.0\%, and 40.2\% across three different scenarios (see Supplementary Materials Fig.~4). Despite varying degrees of improvement, the accuracy consistently increases, indicating that temperature decomposition enables more effective information extraction. 
Specifically, $T_{HF}$ primarily captures thermal effects from cell cycling, while $T_{LF}$ mainly represents ambient temperature fluctuations (see Supplementary Materials Fig.~3). 
Decomposing these signals allows the model to process distinct types of information more effectively. The synergy between these components leads to improved accuracy, as illustrated by the compensation of low-frequency variations in expansion by $T_{LF}$ (see Supplementary Materials Fig.~5). While it may be possible for machine learning models to achieve similar results using surface temperature alone, doing so would require more complex models and long-term datasets. By integrating domain knowledge, we achieve an accurate and data-efficient solution (more analysis about this temperature decomposition could be found in Supplementary Materials Figs. 6-7). 

To examine the contributions of different signals during charging and discharging phases, we propose a novel sensitivity index $\varphi$ (see \nameref{sec:Methods}). This index enables us, for the first time, to make input signals clearly interpretable in relation to battery SOC. 
By leveraging this index, we provide a fundamental understanding of the time-varying importance of different signals and quantify their individual contributions and limitations to estimate SOC. 
As a result, we can identify the ``accuracy barrier'' in SOC estimation for any set of input signals and explain how and why the inclusion of additional signals can enhance accuracy. As illustrated in Fig.~\ref{fig:1_expansionMajor}\textbf{d-h}, $\varphi$ reveals the relative importance of different signals across five scenarios, each representing a distinct signal combination, i.e., \textit{VI}, \textit{E\textbf{T}}, \textit{VI\textbf{T}}, \textit{VIE}, and \textit{VIE\textbf{T}}, corresponding to the violin plots in Fig.~\ref{fig:1_expansionMajor}\textbf{c}. In general, voltage consistently exhibits a high $\varphi$-value in scenarios that include it, indicating its critical role in SOC estimation due to its ability to reflect SOC. Additionally, our analysis of the expansion and temperature signals (Fig.~\ref{fig:1_expansionMajor}\textbf{e}) reveals that expansion has a more pronounced impact on SOC estimation compared to temperature.

\begin{figure}[ht]
    \centering
    \includegraphics[trim=0cm 5cm 5.2cm 0cm, clip, width=0.9\textwidth]{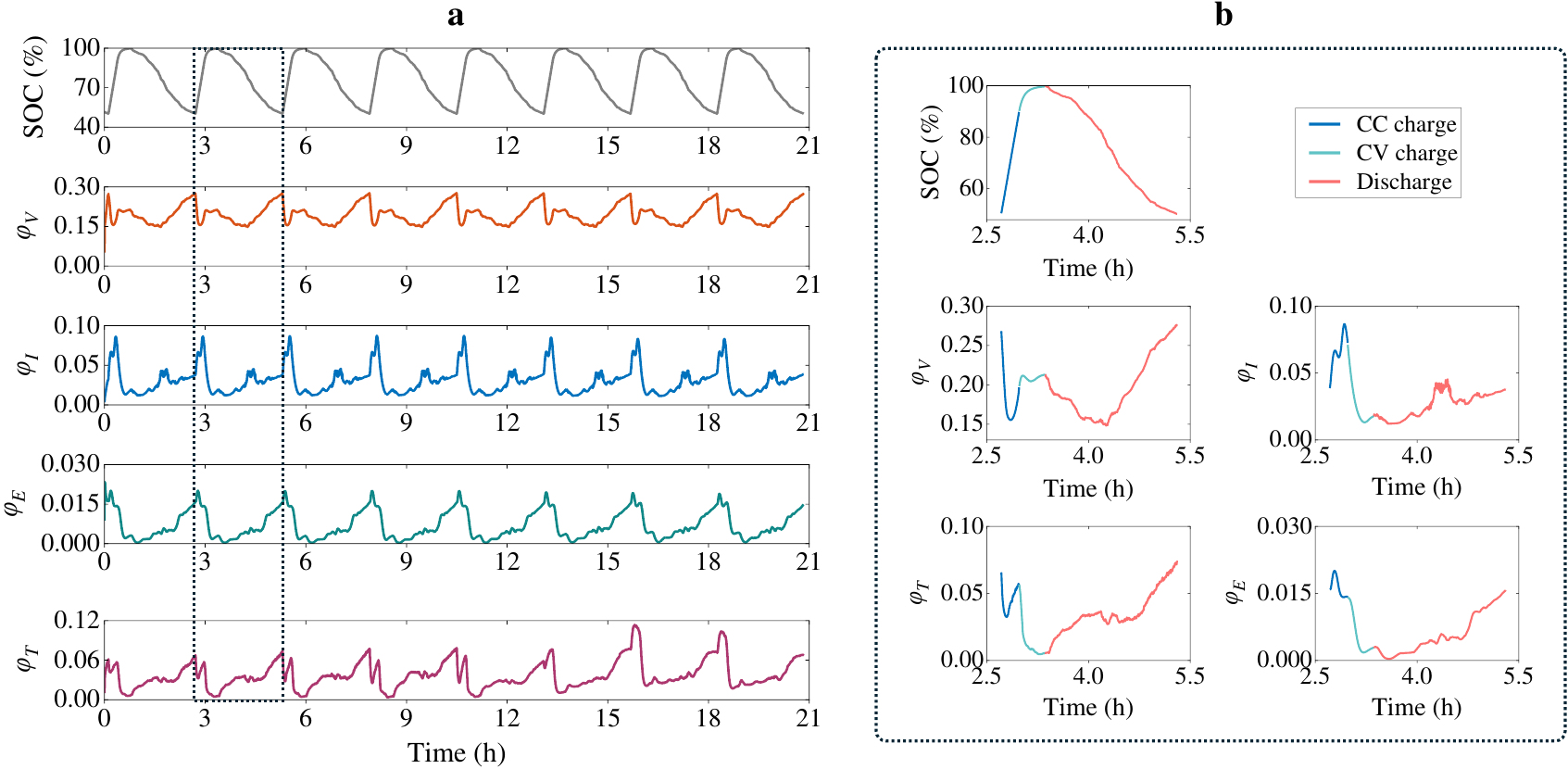}
    \caption{\textbf{Sensitivity analysis of voltage, current, expansion, and temperature signals.}  \textbf{a} Changes in SOC and four sensitivity indices ($\varphi_V$, $\varphi_I$, $\varphi_E$, $\varphi_T$) during charging-discharging cycles. These indices reflect the impact of the corresponding signals—voltage, current, expansion, and temperature—on the accuracy of SOC estimation. \textbf{b} Detailed results from the black dashed frame in \textbf{a}, illustrating changes in the four sensitivity indices and SOC within a single cycle, including constant-current (CC) charge, constant-voltage (CV) charge, and drive cycle discharge. Note that \textit{T} refers to the original cell surface temperature. For complete results including the high-frequency component ($T_{HF}$) and low-frequency component ($T_{LF}$), refer to Supplementary Materials Figs.~9--10.}
    \label{fig:1_1_2VIET_sensitivity}
\end{figure}

To demonstrate the novel time-varying sensitivity index, we present an example in Fig.~\ref{fig:1_1_2VIET_sensitivity}, where all signal types are incorporated. It highlights the importance of the voltage signal, as indicated by its consistently high $\varphi$-values across different conditions. Specifically, during the initial phase of the constant-current (CC) charging process, the voltage signal is crucial as it provides an initial estimation of the SOC. However, as the process progresses, the importance of voltage decreases while the influence of the current signal grows, likely due to the ease with which coulomb counting can be learned by machine learning models during CC charging. 
During the dynamic discharge process, the contribution of the voltage decreases in the mid-part, probably due to the voltage plateau where changes in voltage are small. Concurrently, an increase in $\varphi_I$, indicative of the growing influence of the current signal, is observed. Generally, the current and voltage signals exhibit complementary 
behavior; when the sensitivity of one decreases, the other tends to increase, thereby maintaining high model accuracy. 

Regarding the expansion signal, its primary contribution is evident during the CC charging process, where its sensitivity index $\varphi_E$ increases from an already high baseline, effectively complementing the voltage index (i.e., $\varphi_V$). Additionally, it remains significant towards the end of the discharge process. The surface temperature mainly contributes at the onset of the CC charge and during the mid and final stages of discharge. These findings highlight that the integration of temperature and expansion signals synergistically enhances the performance of the machine learning model with remarkably increased accuracy (improved by 72.2\%).

The results presented above are obtained with regular SOC calibration to mitigate the effects of battery aging. SOC calibration is straightforward in a controlled laboratory setting, where it is well-established that SOC reaches 100\% at the end of CC-CV charging and can be fully discharged to a predetermined lower voltage. However, in practice, regular SOC calibration is costly and typically requires service center intervention, while battery aging gradually affects SOC profiles (see Supplementary Materials Fig.~11). To also explore the performance of different sensors under uncalibrated conditions, we repeat the procedure, and the findings are presented in Supplementary Materials Figs.~12--15. First, we find that new signals become more critical in the absence of SOC calibration. For instance, incorporating temperature and expansion improves accuracy by 90.1\% compared to using only \textit{V-I} signals. Second, the expansion signal gains significant importance, surpassing the current signal in $\varphi$ (see Supplementary Materials Figs.~13 and 15). This is likely due to the expansion signal exhibiting an irreversible increase (Supplementary Materials Fig.~5\textbf{a}), which aids the model in accounting for the aging effects on SOC profiles, thereby enhancing accuracy. Third, the expansion's contribution is notably larger than that of the temperature, as indicated by the sensitivity index (Supplementary Materials Fig.~15). It indicates that the expansion plays a primary role throughout the entire dynamic discharge process, particularly when both voltage and current are at low $\varphi$-values. In contrast, the contribution of temperature is confined to a narrower range within the dynamic discharge process. As a result, as shown in Supplementary Materials Fig.~12, the \textit{V-I-E} fusion yields substantially higher accuracy compared to the \textit{V-I-\textbf{T}} combination. This serves as an example of how the sensitivity index can be used to interpret the outcomes of machine learning models. Overall, regardless of whether SOC calibration is performed, integrating the additional sensors consistently and significantly enhances accuracy.  
Additional validation is conducted under different operating conditions, including low temperatures and high C-rate cycling, as shown in Figs.~16–19 of the Supplementary Materials. The results further confirm the robustness and broad applicability of the proposed multi-sensor fusion approach.

\subsection{\textbf{Cell internal optical signals}}

Fiber optic evanescent wave sensors represent a novel approach for monitoring chemical and electrochemical reactions within batteries \cite{hedman2020fibre, hedman2021fiber, hedman2023fiber}. In this study, we explore, for the first time, the application of such sensors to estimate the SOC in batteries. The sensor is embedded within a cell and placed in direct contact with the cathode (see \nameref{sec:Methods}). It measures both light intensity ($\mathit{\Phi}$) and peak wavelength ($\lambda$), with the signal profiles shown in Fig.~\ref{fig:2_opticalMajor}\textbf{b} (see Supplementary Materials Fig.~27 for additional details). Our results show that incorporating optical signals obviously enhances SOC estimation accuracy. Specifically, Fig.~\ref{fig:2_opticalMajor}\textbf{c} shows that using $\mathit{\Phi}$ and $\lambda$ reduces the MAE to 0.31\%, compared with 0.58\% when relying only on current and voltage measurements—a 46.1\% improvement in accuracy (see Supplementary Materials Fig.~29 for a time-resolved analysis). 
In addition, even without \textit{I-V} signals, optical signals alone achieve an MAE of 0.61\%, indicating their correlation with the battery’s internal state (more information can be found in the Section `Fiber-optic evanescent wave sensor' of the Supplementary Materials).
Beyond this, the optic fiber sensor may be used for other purposes in battery research and development, due to its ability to provide insights into the diffusion and intercalation processes inside a battery cell. 
Potential future applications include enhanced aging monitoring, internal fault detection, thermal management, and improvements in fast-charging technology.

\begin{figure}[ht]
    \centering
    \includegraphics[trim=0cm 2cm 0.3cm 0cm, clip, width=1\textwidth]{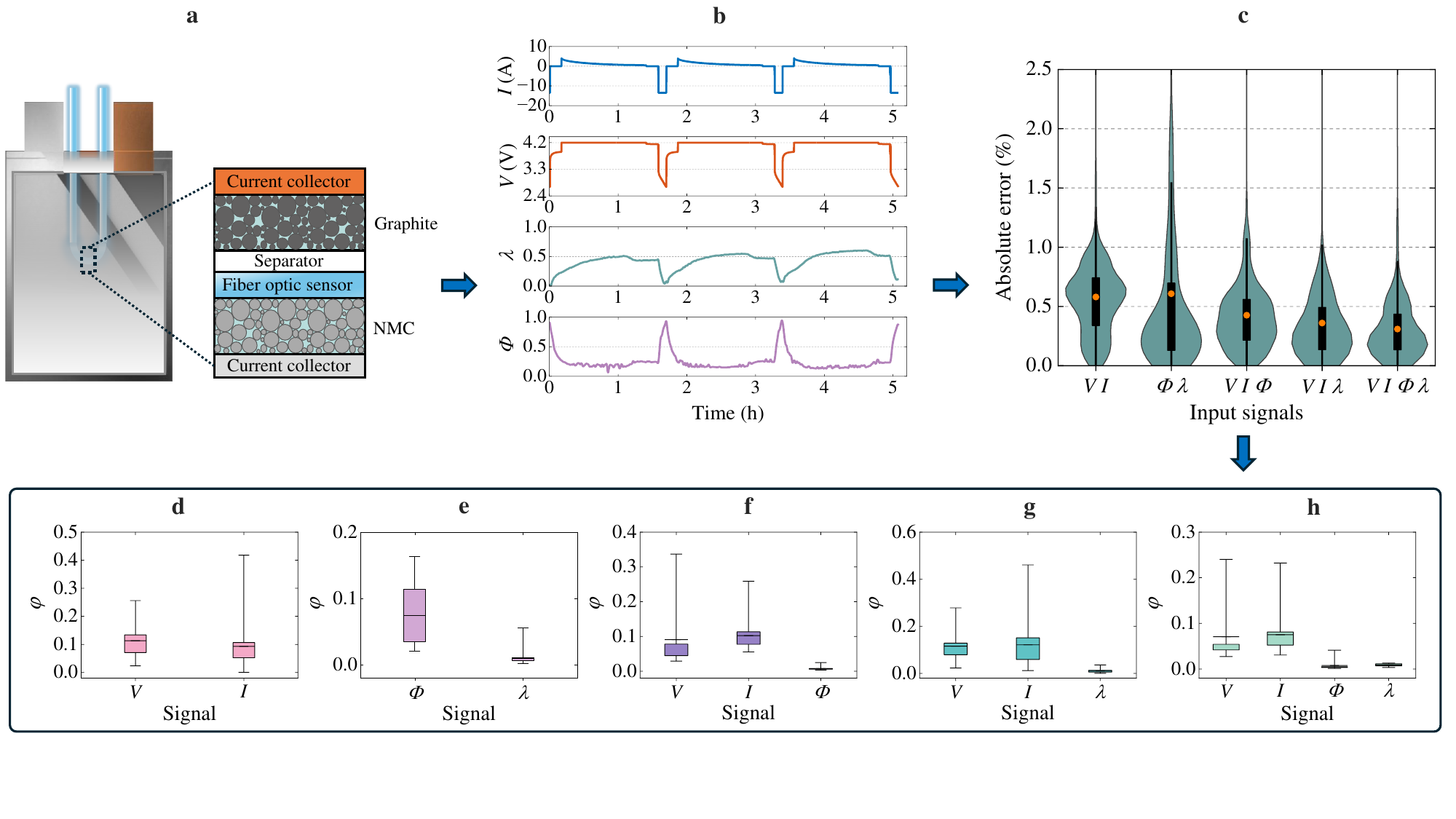}
    \caption{\textbf{Schematic representation and major results for the cell with fiber optic sensor.} The studied signals include  light intensity ($\mathit{\Phi}$) and light peak wavelength ($\lambda$), and conventional measurements, i.e., voltage (\textit{V}) and current (\textit{I}). \textbf{a} Illustration of the test cell. \textbf{b} Three cycles of recorded signal data. \textbf{c} SOC estimation results using different signal combinations. \textbf{d-h} Sensitivity analysis highlighting the contributions of different signals across five scenarios. }
    \label{fig:2_opticalMajor}
\end{figure}

The contribution of different signals is quantified through the proposed sensitivity index $\varphi$. As shown in Fig.~\ref{fig:2_opticalMajor}\textbf{d-h}, the index highlights the significance of different signals across five scenarios, representing the signal combinations $VI$, $\mathit{\Phi}\lambda$, $VI\mathit{\Phi}$, $VI\lambda$, and $VI\mathit{\Phi}\lambda$, each corresponding to one of the violins in Fig.~\ref{fig:2_opticalMajor}\textbf{c}. In general, voltage and current are identified as the most critical signals according to their high $\varphi$-values. Although the optical signals exhibit lower sensitivity index values, they provide crucial supplementary information about material phase changes and solution concentrations \cite{hedman2020fibre, hedman2021fiber}, which are inaccessible through voltage and current measurements. Hence, the inclusion of these optical signals significantly enhances the accuracy.

One noteworthy finding is the increased importance of the current signal and the decreased importance of the voltage signal compared to the results shown in Fig.~\ref{fig:1_expansionMajor}\textbf{d}, which involved a displacement sensor. This change can be attributed to the use of a constant-voltage (CV) charging mode, where the voltage remains constant while the SOC continuously increases (see Supplementary Materials Fig.~28). In this mode, the voltage signal provides limited information about SOC, resulting in its reduced contribution. To show more details, the evolution of the sensitivity index throughout the cycling is given in Supplementary Materials Figs.~30--31. 
During the CV charging phase, $\varphi_V$ is significantly reduced, while the current signal gains prominence, though only at the onset of the CV phase. 
This indicates the presence of an ``accuracy barrier'', where neither the voltage nor the current provides adequate information for accurate SOC estimation during CV charging, as both $\varphi_V$ and $\varphi_I$ exhibit low values.
This aligns with the SOC estimation results over time (Supplementary Materials Fig.~29\textbf{c}), where higher errors are observed during the CV phases.
In addition, during the CC discharge phase, the importance of the current signal is evident, consistent with the findings in Fig.~\ref{fig:1_1_2VIET_sensitivity} on its relevance in CC processes.

In terms of optical signals, the intensity is particularly significant during the CC discharge and rest periods, where its influence overlaps with that of current and voltage (see Supplementary Materials Fig.~31). In contrast, the peak wavelength is distinctive, as its primary contribution occurs during the CV charging phase, making it essential to compensate for the limited information from the other three signals. This explains why the combination of voltage, current, and peak wavelength ($V$-$I$-$\lambda$) achieves higher accuracy than the combination using intensity ($V$-$I$-$\mathit{\Phi}$), as shown in Fig.~\ref{fig:2_opticalMajor}\textbf{c}. 

\subsection{\textbf{Cell force and electrode potential signals} }

\begin{figure}[ht]
    \centering
    \includegraphics[trim=0cm 0.9cm 7cm 0cm, clip, width=0.8\textwidth]{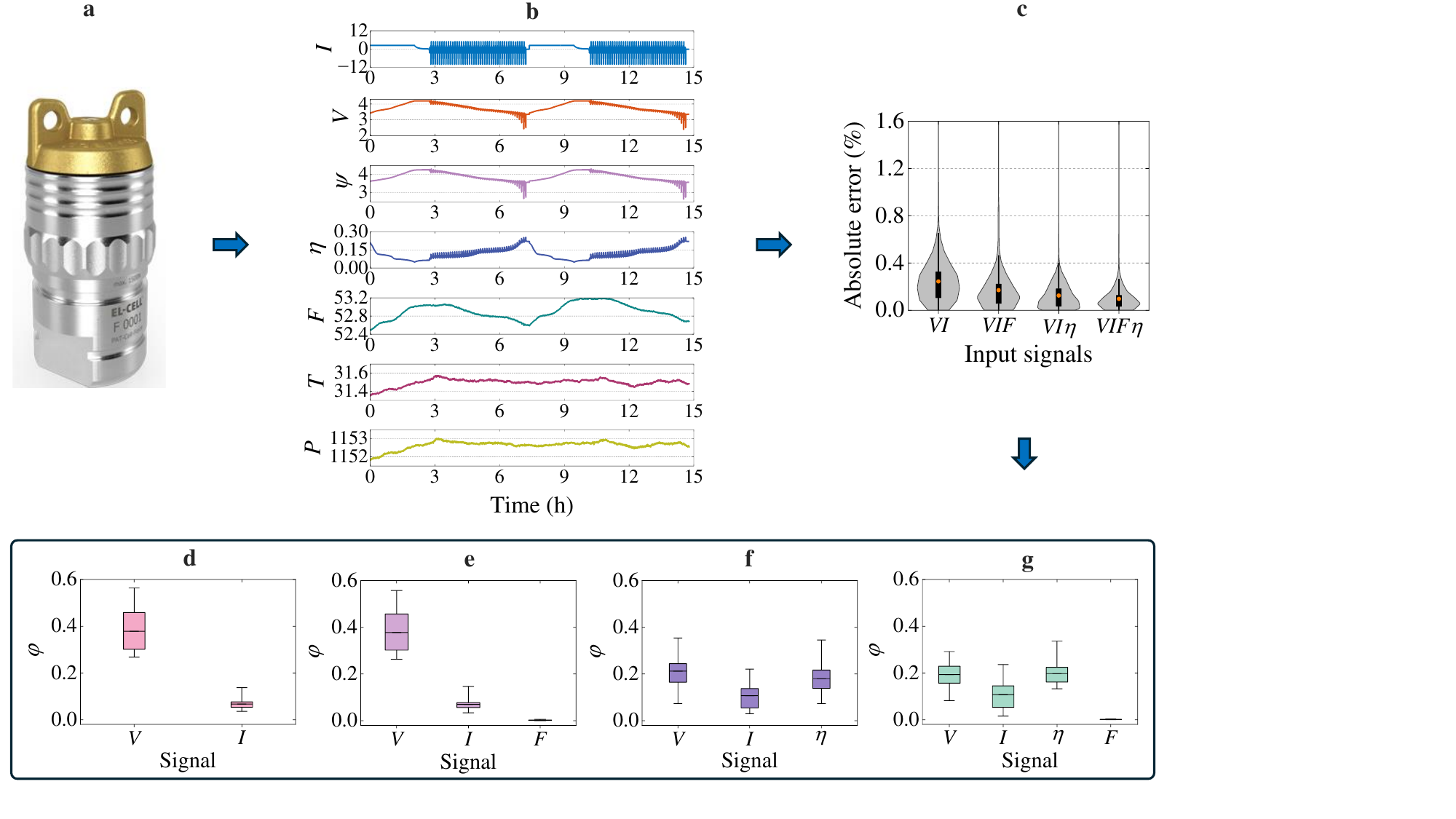}
    \caption{\textbf{Schematic representation and major results for the cell with multiple sensors (PAT-Cell-Force from EL-CELL).} The studied signals include cathode potential ($\mathit{\psi}$), anode potential ($\mathit{\eta}$), force ($\mathit{F}$), internal temperature ($T_{\mathrm{in}}$), gas pressure ($\mathit{P}$), and conventional measurements, i.e., voltage (\textit{V}) and current (\textit{I}). \textbf{a} Illustration of the test cell. \textbf{b} Two cycles of recorded signal data. \textbf{c} The SOC estimation results for different signal combinations, with each combination including voltage, current, and one additional signal. \textbf{d} SOC estimation results using the signals of $V$, $I$, $\mathit{\eta}$, and $\mathit{F}$. \textbf{e-h} Sensitivity analysis highlighting the contributions of different signals across four scenarios.}
    \label{fig:4_ElcellMajor}
\end{figure}

We also investigate the use of electrode potentials, force, internal temperature, and gas pressure, in estimating the SOC of a Li-ion battery. These signals are obtained using a test cell equipped with sensors for current, voltage, force, temperature, and gas pressure, as shown in  Fig.~\ref{fig:4_ElcellMajor}\textbf{a}. The signal profiles are presented in Fig.~\ref{fig:4_ElcellMajor}\textbf{b} (see Supplementary Materials Fig.~32 for further details). This study explores different combinations of these signals to optimize SOC estimation. Initially, we assess the impact of incorporating novel signals into traditional measurements (i.e., $V$ and $I$). The results show that adding anode potential and force signals substantially improves the accuracy. Specifically, the inclusion of these two signals reduced the MAE from 0.24\% to 0.10\%, increasing the accuracy by 60.6\%, as shown in Fig.~\ref{fig:4_ElcellMajor}\textbf{c} (see Supplementary Materials Figs.~33--34 for the error trajectories).

The contributions of all involved signals are quantified by the proposed sensitivity index ($\varphi$), as shown in Fig.~\ref{fig:4_ElcellMajor}\textbf{d-g}. It indicates the importance of different signals across four scenarios, i.e., the signal combinations $VI$, $VIF$, $VI\mathit{\eta}$, and $VI\mathit{\eta}F$, each corresponding to one of the violins in Fig.~\ref{fig:4_ElcellMajor}\textbf{c}.
Voltage and anode potential emerge as the most critical signals, with high $\varphi$-values. This is attributed to their direct correlation with the SOC, governed by the stable relationship between the open circuit potentials and SOC. In contrast, although the force signal exhibits lower $\varphi$-values, it provides auxiliary information mainly into the anode material intercalation and de-intercalation processes, phenomena that the electrical signals alone cannot well characterize. Therefore, including the force signal further enhances estimation accuracy: the MAE decreases from 0.12\% to 0.10\% when changing from the $VI\eta$ combination to the $VI\eta F$, representing an accuracy improvement of 22.9\% (Fig.~\ref{fig:4_ElcellMajor}\textbf{c}).

Another key finding is that the contributions of voltage and current exhibit high similarity to the results obtained from the cell with a displacement sensor (compare Fig.~\ref{fig:4_ElcellMajor}\textbf{e} with Fig.~\ref{fig:1_expansionMajor}\textbf{d}). This can be attributed to the similar battery chemistry and test procedure, specifically the CC-CV charging followed by dynamic discharge. This consistency underscores the reliability of the proposed index in interpreting signal contributions. When examining this index over time (see Fig.~\ref{fig:1_1_2VIET_sensitivity} and Supplementary Materials Fig.~35), its evolution patterns remain consistent. For instance, the $\varphi_V$ value shows a V-shaped pattern during both the CC charge and dynamic discharge processes in both cells. The reciprocal relationship between current and voltage signals is evident, explaining their combined effectiveness in SOC estimation. 

The anode potential signal primarily influences the later stages of CC charging and dynamic discharge, effectively complementing the current signal  (see Supplementary Materials Fig.~36). Notably, the anode potential’s contribution closely resembles that of the voltage, as $\varphi_{\eta}$ is similar to $\varphi_V$ across different cycling stages. This can be explained by the relationship \(V = \psi - \eta\). However, clear differences emerge during specific phases, such as the end of dynamic discharge and the onset of CV charging, where the anode potential contributes more than the voltage. This indicates that anode potential provides unique insights not fully captured by voltage, even though voltage contains much of the information related to anode potential. Moreover, the force signal offers distinctive contributions during periods when both voltage and current show reduced $\varphi$-values, while the force signal's $\varphi$ increases. These instances occur at the end of dynamic discharge, the start and middle of CC charging, and the middle of dynamic discharge. Together, the anode potential and force signals help overcome the ``accuracy barrier'' by providing complementary information under certain conditions. Consequently, integrating these signals significantly improves the model’s performance, boosting accuracy by 60.6\%.

\subsection{\textbf{Discussion of feasibility and applicability}}

We evaluate eleven types of signals in this study to assess their potential for improving SOC estimation for Li-ion batteries. Our results demonstrate that integrating additional signals can significantly enhance SOC estimation accuracy. However, the objective of this work is not to replace conventional SOC estimation signals (i.e., current and voltage), which remain highly effective, but rather to complement them, enhancing overall accuracy and reliability.

Implementing multi-sensor integration in real-world BMS poses several challenges. Not all signal types provide equal contributions to improving SOC estimation. For instance, our findings indicate that cathode potential, internal temperature, and gas pressure offer minimal improvements in SOC accuracy. The minimal impact of internal temperature observed in this study could be attributed to the small size and low heat generation in coin-type test cells. Nonetheless, these measurements might be highly beneficial for other critical purposes, such as battery aging estimation and fault monitoring.

Signals related to cell expansion, surface temperature, light intensity, wavelength, anode potential, and mechanical force substantially enhance SOC accuracy when integrated with current and voltage. Among these, surface temperature is already routinely monitored at the module or pack level, making it the most accessible additional measurement for immediate integration. 
Cell expansion and mechanical force measurements also have substantial potential, given their ease of implementation through inexpensive and compact sensors. 
Nevertheless, real-time computational requirements and signal synchronization remain practical challenges for multi-sensor systems, though continued advancements in computational technology should mitigate these issues in the near future.
Particularly, this study demonstrates that employing an LSTM network for SOC estimation incurs only a minimal increase in computational cost when integrating multiple sensor signals. For example, in the case described in Section~\ref{sec:cell-expansion}, the elapsed computational time per step increased marginally from $5.98\,\mu\text{s}$ for the \textit{V-I} inputs to $6.01\,\mu\text{s}$ when including cell expansion and temperature data (\textit{VIE\textbf{T}} inputs).

Regarding optical signals, specifically the light intensity and wavelength measured by fiber-optic sensors, we recognize their primary value currently being within research. Their implementation in commercial battery cells is complicated by invasive installation requirements, as optical fibers must directly penetrate the cell and interface with electrode materials. 
Additionally, the instrumentation, including specialized light sources and spectrometers, still tends to be far too expensive for widespread commercial deployment. Nonetheless, fiber-optic sensing offers exceptional capabilities for research purposes by providing valuable insights into internal material properties and chemical or electrochemical reaction mechanisms \cite{hedman2020fibre, hedman2021fiber, hedman2023fiber}.

Beyond the above sensing methods, electrochemical impedance spectroscopy (EIS) and acoustic sensing techniques also offer promising capabilities for battery SOC estimation. EIS is a robust diagnostic tool particularly valuable for battery SOH evaluation, as it provides insights into internal degradation mechanisms and associated patterns~\cite{jones2022impedance, zhang2020identifying}. 
For example, EIS-enabled AFE chips from Analog Devices can measure a battery’s impedance, allowing for the detection of battery degradation~\cite{bautista2024helpstat}.
For SOC estimation, EIS primarily contributes by supplying critical aging information, aiding in the calibration of battery models. 
This calibration, in turn, enhances the accuracy and robustness of SOC estimation, particularly in aged cells~\cite{li2024soh, wang2023application}.

Acoustic sensing has also emerged as a promising alternative for estimating battery SOC by exploiting the mechano-electrochemical interactions within Li-ion cells. 
During charge and discharge cycles, internal battery properties, including elastic modulus, density, and structural dimensions, undergo changes that significantly affect acoustic wave propagation~\cite{galiounas2022battery}. Parameters such as the time of flight and signal amplitude of acoustic waves can reflect these mechanical properties~\cite{ladpli2018estimating, wang2023active}. 
In this study, we also investigate the integration of acoustic signals for battery SOC estimation. Our findings demonstrate an obvious improvement in SOC estimation accuracy when acoustic data is incorporated. This aligns well with our previous observations using mechanical, thermal, and optical signals, confirming the robustness and effectiveness of our proposed multi-sensor approach. The detailed results can be found in the Supplementary Materials (see Figs.~43--45). For its real application, as an example, Volvo Trucks has developed a prototype acoustic early-warning system that detects infrasound generated by gas bubbles forming within a battery cell, a subtle early indicator of thermal runaway, enabling intervention before temperatures rise dangerously~\cite{movaghar2025passive}.

\section{Conclusions}
While SOC reflects the coupled multiphysics within battery cells, existing BMS heavily relying on current and voltage measurements alone is fundamentally limited in accurately tracking SOC evolution. 
To overcome these limitations, we have developed an explainable machine-learning approach to integrate novel mechanical, thermal, gas, optical, and/or additional electrical sensors with traditional measurements. This approach is capable of physically interpreting and quantitatively evaluating the dynamic contributions of individual input signals, thereby providing a theoretical foundation for optimized sensor deployment and utilization. 

Extensively trained and tested on three unique datasets comprising eleven different signals, this multi-sensor fusion approach demonstrated a transformative enhancement in SOC estimation accuracy across Li-ion batteries of diverse chemistries and operating conditions. Specifically, we achieved improvements of 46.1\% to 74.5\% over traditional methods based solely on voltage and current. 
As a result, the SOC operating window, which has previously been constrained overly conservatively, can now be safely and more accurately extended, yielding substantial economic and environmental benefits for industries along the battery value chain. 
Beyond SOC estimation, this machine learning sensor fusion approach holds promise for advancing the monitoring of other critical battery states, ultimately promoting safer and more efficient battery usage. Future experimental investigations will further explore its applicability and validate the broader potential benefits.

\section{Methods}
\label{sec:Methods}

\subsection{\textbf{Experimental and data generation}}

Three different types of cells were used in this work. 
They are laboratory-fabricated NMC/graphite pouch cells \cite{mohtat2022comparison}, commercial AMTE NMC/graphite pouch cells, and NMC/graphite coin cells. 
For each cell type, different sensors were employed to obtain in total eleven different signals, including battery expansion, force, gas pressure, internal and surface temperatures, light intensity and peak wavelength, anode and cathode potentials, current, and voltage. The specifications of these cells are summarized in Supplementary Materials Table 1. Testing these cells yielded three comprehensive datasets.

Each of the laboratory-fabricated NMC/graphite pouch cells was mounted in a fixture (see Fig.~\ref{fig:1_expansionMajor}\textbf{a}) and placed in a climate chamber (Cincinnati Ind., USA), where cell expansion was monitored using a displacement sensor (Keyence, Japan). 
The climate chamber controls the cell's ambient temperature during cycling, with temperature readings taken via a K-type thermocouple (Omega, USA) positioned on the cell's surface. 
Initially, the cell was charged at a constant current (CC) of 1.5C, corresponding to 7.5 A, until the voltage reaches 4.2 V. 
This CC phase was followed by constant voltage (CV) charging until the current fell below C/50. 
Subsequently, discharge commenced until the depth of discharge (DoD) reached 50\%, following a synthetic drive cycle to mimic electric vehicle (EV) operation  
(depicted in Supplementary Materials Fig.~2). Further details on the conducted tests can be found in reference  \cite{mohtat2022comparison}.

The AMTE NMC/graphite pouch cell was assembled with a fiber optic sensor embedded between the separator and cathode  
(see Fig.~\ref{fig:2_opticalMajor}\textbf{a}), and was placed in a room maintained at approximately 25°C. 
In this configuration, the sensor's surface was in direct contact with the cell's cathode material and surrounded by electrolyte. 
During cycling, the cathode material composition and electrolyte concentration evolved, altering the absorption of specific light wavelengths. 
This resulted in variations in both the detected light intensity and peak wavelength, thereby providing new insights into previously unavailable internal information. 
Light signals were monitored using a fiber optic sensor (Insplorion AB, Sweden), while battery cycling was performed using a battery cycler (Biologic, France). 
The test data used to estimate the SOC comprised cycles involving constant voltage charging, constant current discharge, and rest periods. 
During CV charging, a voltage of 4.2V was maintained until the current decreased to 1/20 C (0.45 A), followed by a CC discharge at 1.5 C until the voltage reached 2.7V, and a rest period of 10 minutes concluding each cycle.

The NMC/graphite coin cell was assembled using a PAT-Cell-Force setup (El-Cell, Germany), equipped with sensors for current, voltage, force, temperature, and gas pressure. 
This configuration enabled seven different measurements, as shown in Fig.~\ref{fig:4_ElcellMajor}\textbf{b}. 
In addition, the cell channel itself functioned as a tester to control the applied current. 
Testing was conducted at room temperature of approximately 25°C. 
The obtained data included cycles involving CC charging, CV charging, and dynamic discharge processes, specifically the dynamic stress test (DST). 
During CC charging, a current of 2.5 mA (0.5 C) was applied until the voltage reached 4.2 V. 
This was followed by CV charging at 4.2 V until the current decreased to 0.1 mA (C/50). 
A 10-minute rest period was then applied. 
Subsequently, the DST was implemented until the voltage dropped to 2.5 V, completing one cycle.

The observed changes in sensor signals directly reflect battery structural evolution and internal electrochemical processes. For example, Li-ion batteries expand/contract during cycling due to Li-ion intercalation/deintercalation within electrode materials. Upon charging, the graphite anode expands (around 13\% by volume), while the cathode undergoes minor contraction (e.g., around 3\% for NMC cathodes, and 1\% for lithium cobalt oxide cathodes) \cite{huang2022onboard}. Given the anode's dominant volume change, full-cell mechanical responses primarily reflect electrochemical processes occurring at the graphite anode. When the battery cell is mechanically constrained or fixed, these structural changes are translated into measurable variations in mechanical force. 
In addition, fiber-optic sensors detect internal battery state changes through the interaction of evanescent waves with the electrode materials (more information can be found in the Section `Fiber-optic evanescent wave sensor' of the Supplementary Materials).
For these sensors, the input light with constant intensity and wavelength interacts with the surrounding battery materials; changes in material composition or structure alter the evanescent wave absorption, thereby modulating the output optical signals. Consequently, Li-ion  cycling affects both the intensity and peak wavelength of the fiber-optic sensor's outputs, facilitating monitoring of internal electrode conditions.

\subsection{\textbf{Explainable machine-learning model} }

{\bf (1) Selection and formulation of the base model.} Recurrent neural networks (RNNs) are a specialized class of artificial neural networks designed to recognize patterns in sequential data. They are applicable to a variety of time-series tasks, such as speech recognition, natural language understanding, and machine translation \cite{sherstinsky2020fundamentals}. Despite their versatility, classical RNNs struggle with long-range dependencies due to vanishing or exploding gradients—issues that compromise their ability to learn from extended sequences \cite{chemali2017long}. Long short-term memory networks (LSTMs) were introduced to address these limitations and have since become widely used \cite{chemali2017long, yu2019review}. 
LSTMs enhance performance through a gating mechanism that regulates information flow, selectively retaining or discarding data to preserve critical information over prolonged intervals. 
This capability makes LSTMs particularly valuable for complex tasks involving extensive contextual data, such as time series estimation and predictions such as those in this study. Moreover, LSTMs excel in capturing the nonlinear dynamics of complex systems without requiring feature construction or engineering \cite{chemali2017long}. In this work, we leverage LSTMs to conduct a fair and systematic comparison of multiple sensors and their combinations in SOC estimation under diverse operating conditions. This approach aims to provide robust, feature-independent insights that can inform sensor selection and enhance SOC estimation accuracy.

In our machine learning models, the input sequence comprises data from different sensors, e.g., measurements of cell expansion, light intensity, voltage, and current, while the SOC values form the output sequence. 
These labeled input-output data samples for training the LSTM models are mathematically defined within the dataset $D$
    \begin{equation}
    D = \{ \left(\Psi_1, y_1\right), (\Psi_2, y_2),
    \ldots, (\Psi_k, y_k), \ldots, (\Psi_M, y_M) \},
    \end{equation}
where \(y_k\) represents the SOC calibrated at time step \(k\) in a well-controlled laboratory environment with high-resolution sensors, and it plays the ground truth. $M$ is the size of the training set. \(\Psi_k\) is the corresponding vector of inputs, defined as $\Psi_k=[\xi_1(k), \ldots, \xi_l(k), \ldots, \xi_L(k)]$, where \(\xi_l(k)\) is the \(l\)-th signal measured from the battery at $k$, and $L$ is the number of signals. 
We choose a LSTM model with two layers, each with 100 hidden units. 
The operations of each hidden unit with the input, forget, and output gates can be formulated as \cite{chemali2017long, yang2020state}
\begin{align}
    & \text{Input gate: }
    \quad\quad\quad i_k  = \sigma\left(W_{\Psi_i} \Psi_k + W_{y_i} \hat{y}_{k-1} + b_i\right), \label{Eq:Model1}\\
    & \text{Forget gate: }
    \:\:\quad\quad f_k = \sigma\left(W_{\Psi_f} \Psi_k + W_{y_f} \hat{y}_{k-1} + b_f\right), \label{Eq:Model2}\\
    & \text{Cell state update: }
    \tilde{C}_k = \tanh\left(W_{\Psi_c} \Psi_k + W_{y_c} \hat{y}_{k-1} + b_c\right), \label{Eq:Model3}\\
    & \text{Final cell state: }
    \:\:\:\: C_k = f_kC_{k-1} + i_k\tilde{C}_k, \label{Eq:Model4}\\
    & \text{Output gate: }
    \quad\quad O_k = \sigma\left(W_{\Psi_o} \Psi_k + W_{y_o} \hat{y}_{k-1} + b_o\right), \label{Eq:Model5}\\
    & \text{Final output: }
    \:\quad\quad \hat{y}_k = O_k\tanh\left(C_k\right),\label{Eq:Model6}
\end{align}
where \( \sigma \) denotes the sigmoid function, \( \tanh \) is the hyperbolic tangent function, \( W \) and \( b \) represent the weight and bias, respectively, and $\hat{y}$ is the model-predicted output.

\textbf{(2) The explainable model enabled by time-varying sensitivity.} Existing machine learning models are typically evaluated only on their final predictions, leading them to be labeled as ``black boxes'' \cite{beckh2021explainable}. Shapley values, derived from cooperative game theory, are widely used to interpret feature contributions and importance in machine learning predictions \cite{shapley1953value,lundberg2017unified}. 
However, classical Shapley values exhibit limitations when applied to complex models, especially deep learning frameworks handling time-series data \cite{belle2021principles}. 
Specifically, the assumption of feature independence underpinning Shapley values largely restricts their interpretability for time-series data, where features are inherently interdependent and sequentially structured.

\begin{figure}[ht!]
    \centering
    \includegraphics[trim=0cm 0.05cm 0cm 0cm, clip, width=0.78\textwidth]{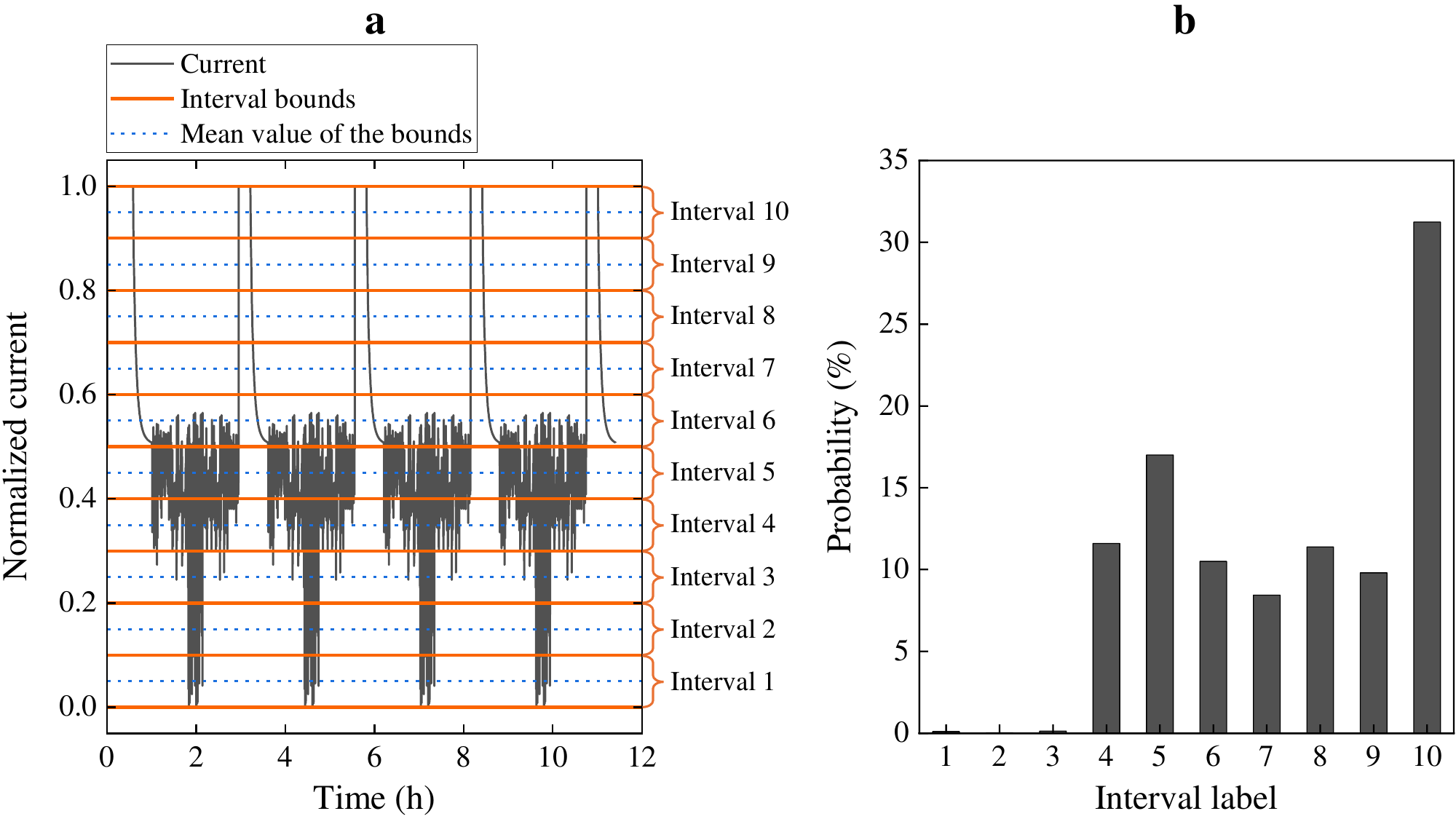}
    \caption{\textbf{An example to illustrate the interval division and probability distribution for calculating the sensitivity index $\varphi_{l}$.} \textbf{a} Schematic showing how to define intervals for the current signal. In this example, the current signal is divided into 10 intervals (i.e., \textit{H} = 10). The orange lines represent the interval bounds, and the light-blue dashed lines indicate the mean values for each interval $h$, i.e., $\bar{\xi}_{l,h}$.  \textbf{b} The probability distribution ($P_{l,h}$) of the current signal across intervals 1 to 10.}
    \label{fig:5_SensitivityIndexParameterExplain1}
\end{figure}

To overcome these limitations, we propose a novel sensitivity index designed explicitly for LSTM-based time-series models, enabling the real-time quantification of signal contributions. The proposed time-varying sensitivity index is defined as:

    \begin{equation}
    \varphi_{l}(k) = \sum_{h=1}^{H} P_{l,h}
    \left|f\left(\Psi_{k-S+1:k}\right) 
    - 
    f\left( 
    \{\Psi_{k-S+1:k}|\xi_{l} = \bar{\xi}_{l,h}\}
    \right)\right|, 
     \label{eq:SensitivityIndex}
    \end{equation}

where $\varphi_{l}(k)$ represents the sensitivity index for the $l$-th signal at time step $k$, $H$ denotes the number of intervals into which the signal is divided, $h$ is the index of the interval, and $f(\cdot)$ is the trained LSTM model specified by Equations \eqref{Eq:Model1}--\eqref{Eq:Model6}. 
The term $\left(\Psi_{k-S+1:k}\right)$ indicates the set of all $L$ input signals within a sliding time window from $k-S+1$ to $k$, with window length $S$.
The term $\{\Psi_{k-S+1:k}|\xi_{l} = \bar{\xi}_{l,h}\}$ represents the input vectors where the signal $\xi_{l}$ is held constant at a value $\bar{\xi}_{l,h}$, derived as the mean within interval $h$. 
The weight $P_{l,h}$ represents the probability associated with selecting the mean value $\bar{\xi}_{l,h}$. Specifically, $P_{l,h}$ is derived from the probability that signal $l$ falls within interval $h$.
A detailed illustration clarifying these parameters is presented in Fig.~\ref{fig:5_SensitivityIndexParameterExplain1}.

This sensitivity index quantifies the local, interval-specific relevance of individual signals while preserving temporal continuity, thus addressing the limitations inherent in traditional Shapley value computations. While $\varphi_{l}(k)$ is intuitively constructed and practically oriented, it aligns closely with established sensitivity analysis principles, evaluating the model's response to controlled, localized input perturbations. To further enhance robustness and validate that $\varphi_{l}(k)$ captures authentic rather than spurious contributions, we conduct perturbation experiments (see Figs.~20-21 in the Supplementary Materials). By perturbing input signals and observing corresponding output variations, the results show how the sensitivity index can reflect the significance of each signal. 

{\bf (3) Training and evaluation.}
Each of the three datasets was split, with 80\% allocated to training and 20\% to testing.
The network is trained using the ‘adam’ optimizer  \cite{kingma2014adam}, which dynamically adjusts learning rates based on the gradient of the loss function. The training process is set to a maximum of 6000 epochs, allowing enough iterations for the network to learn from the data. The training progress was tracked using a graphical display of training metrics in a non-verbose mode, facilitating early detection of potential issues, such as overfitting and gradient exploding.

MAE and RMSE are chosen to evaluate the performance of each machine learning model, defined as 
\begin{align}
    \text{MAE} =\:& \frac{\sum_{i=1}^N |y_i - \hat{y}_i|}{N}\: , \\
    \text{RMSE} =\:& \sqrt{\frac{1}{N} \sum_{i=1}^{N} (y_i - \hat{y}_i)^2}\: ,
    \end{align}
where \(N\) is the number of test samples.

    \section*{Data availability}
All data used in this study are available
\href{https://drive.google.com/drive/folders/1z6uo_XGRykryIokeOnx_Qaf56yRHMEcP?usp=sharing}{(Link)}.

\section*{Code availability}
Code for data processing is available \href{https://drive.google.com/drive/folders/1z6uo_XGRykryIokeOnx_Qaf56yRHMEcP?usp=sharing}{(Link)}. Code for the modeling work is available from the corresponding authors upon request.  

\section*{Declaration of generative AI and AI-assisted technologies in the writing process}
During the preparation of this work, the authors used ChatGPT4o in order to polish the language. After using this tool, the authors reviewed and edited the content as needed and take full responsibility for the content of the publication.

 \bibliographystyle{elsarticle-num}
 \bibliography{reference}

\end{document}